\documentclass[fleqn,10pt]{wlscirep}

\usepackage[utf8]{inputenc}
\usepackage[T1]{fontenc}
\usepackage{multirow}
\title{Answering Mermin's Challenge with Conservation per No Preferred Reference Frame}

\author[1,*]{W.M. Stuckey}
\author[2,3]{Michael Silberstein}
\author[4]{Timothy McDevitt}
\author[5]{T.D. Le}
\affil[1]{Department of Physics, Elizabethtown College, Elizabethtown, PA 17022, USA}
\affil[2]{Department of Philosophy, Elizabethtown College, Elizabethtown, PA 17022, USA}
\affil[3]{Department of Philosophy, University of Maryland, College Park, MD 20742, USA}
\affil[4]{Department of Mathematical Sciences, Elizabethtown College, Elizabethtown, PA 17022, USA}
\affil[5]{ College of Computing,  Georgia Institute of Technology, Atlanta, GA 30332, USA}
\affil[*]{stuckeym@etown.edu}
\begin{document}

\begin{abstract}
In 1981, Mermin published a now famous paper titled, ``Bringing home the atomic world: Quantum mysteries for anybody'' that Feynman called, ``One of the most beautiful papers in physics that I know.'' Therein, he presented the ``Mermin device'' that illustrates the conundrum of quantum entanglement per the Bell spin states for the ``general reader.'' He then challenged the ``physicist reader'' to explain the way the device works ``in terms meaningful to a general reader struggling with the dilemma raised by the device.'' Herein, we show how ``conservation per no preferred reference frame (NPRF)'' answers that challenge. In short, the explicit conservation that obtains for Alice and Bob's Stern-Gerlach spin measurement outcomes in the same reference frame holds only on average in different reference frames, not on a trial-by-trial basis. This conservation is SO(3) invariant in the relevant symmetry plane in real space per the SU(2) invariance of its corresponding Bell spin state in Hilbert space. Since NPRF is also responsible for the postulates of special relativity, and therefore its counterintuitive aspects of time dilation and length contraction, we see that the symmetry group relating non-relativistic quantum mechanics and special relativity via their ``mysteries'' is the restricted Lorentz group.
\end{abstract}

\flushbottom
\maketitle

\thispagestyle{empty}

\section*{Introduction}\label{Secintroduction}

\label{Secintro}
Physics is a science dedicated to understanding the physical world and, as astrophysicist and writer Adam Becker points out \cite[p. 7]{becker}:
\begin{quote}
science is about more than mathematics and predictions -- it's about building a picture of the way nature works. And that picture, that story about the world, informs both the day-to-day practice of science and the future development of scientific theories, not to mention the wider world of human activity outside of science. 
\end{quote}
For example, geocentricism gave way to heliocentricism in part due to the principle of relativity, i.e., the laws of physics are the same in all inertial reference frames, which is sometimes referred to as ``no preferred reference frame'' (NPRF). Newtonian mechanics and special relativity are both based on the principle of relativity. The difference between the Galilean transformations of Newtonian mechanics and the Lorentz transformations of special relativity resides in the fact that the speed of light is finite, so NPRF entails the light postulate of special relativity, i.e., that everyone measure the same speed of light \textit{c}, regardless of their motion relative to the source. If there was only one reference frame for a source in which the speed of light equaled the prediction from Maxwell's equations ($c = \frac{1}{\sqrt{\mu_o\epsilon_o}}$), then that would certainly constitute a preferred reference frame. 

\begin{figure}[h!]
\begin{center}
\includegraphics [height = 55mm]{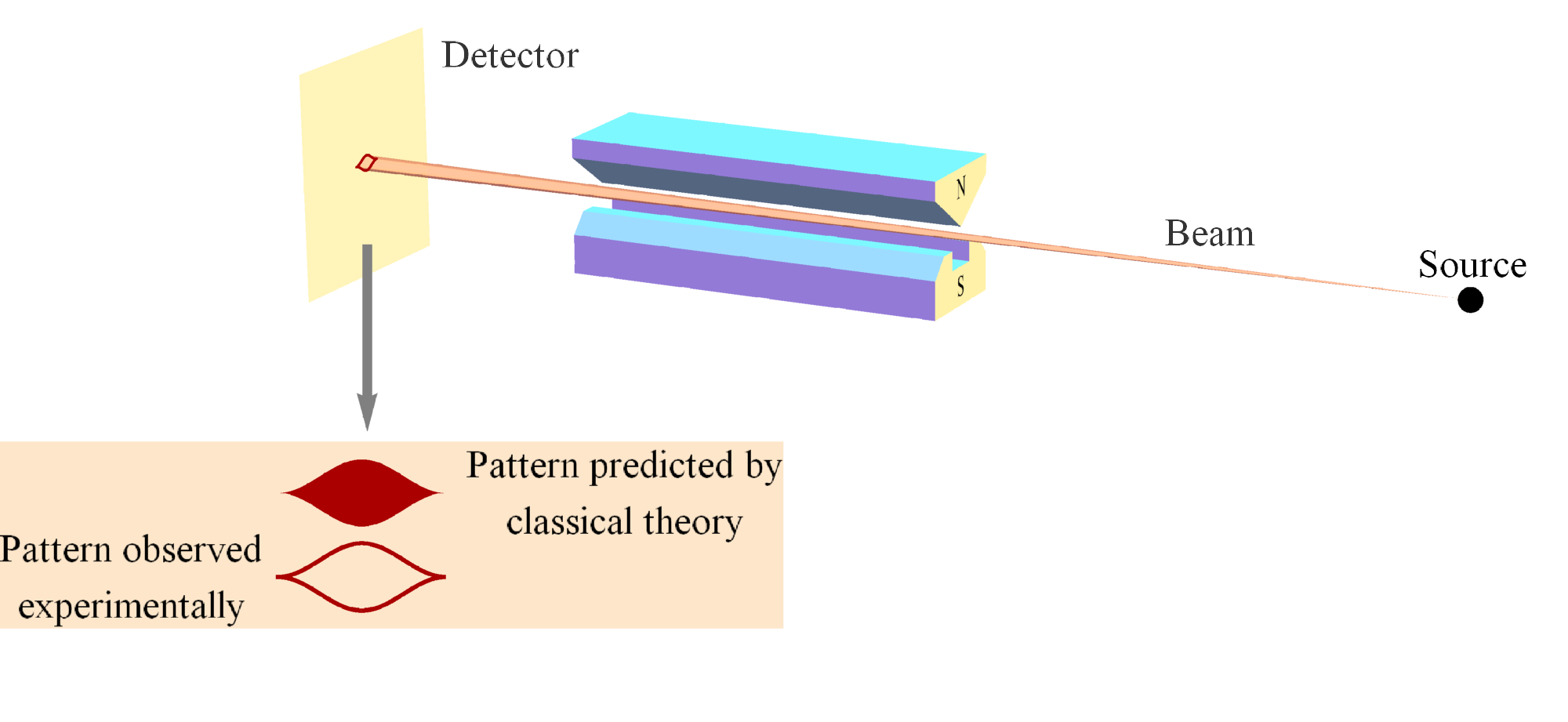}  \caption{A Stern-Gerlach (SG) spin measurement showing the two possible outcomes, up ($+\frac{\hslash}{2}$) and down ($-\frac{\hslash}{2}$) or $+1$ and $-1$, for short. The important point to note here is that the classical analysis predicts all possible deflections, not just the two that are observed. This binary (quantum) outcome reflects Dakic and Brukner's first axiom in their reconstruction of quantum theory, ``An elementary system has the information carrying capacity of at most one bit'' \cite{dakic}. The difference between the classical prediction and the quantum reality uniquely distinguishes the quantum joint distribution from the classical joint distribution for the Bell spin states \cite{garg}.} \label{SGExp}
\end{center}
\end{figure}

\begin{figure}[h!]
\begin{center}
\includegraphics [height = 50mm]{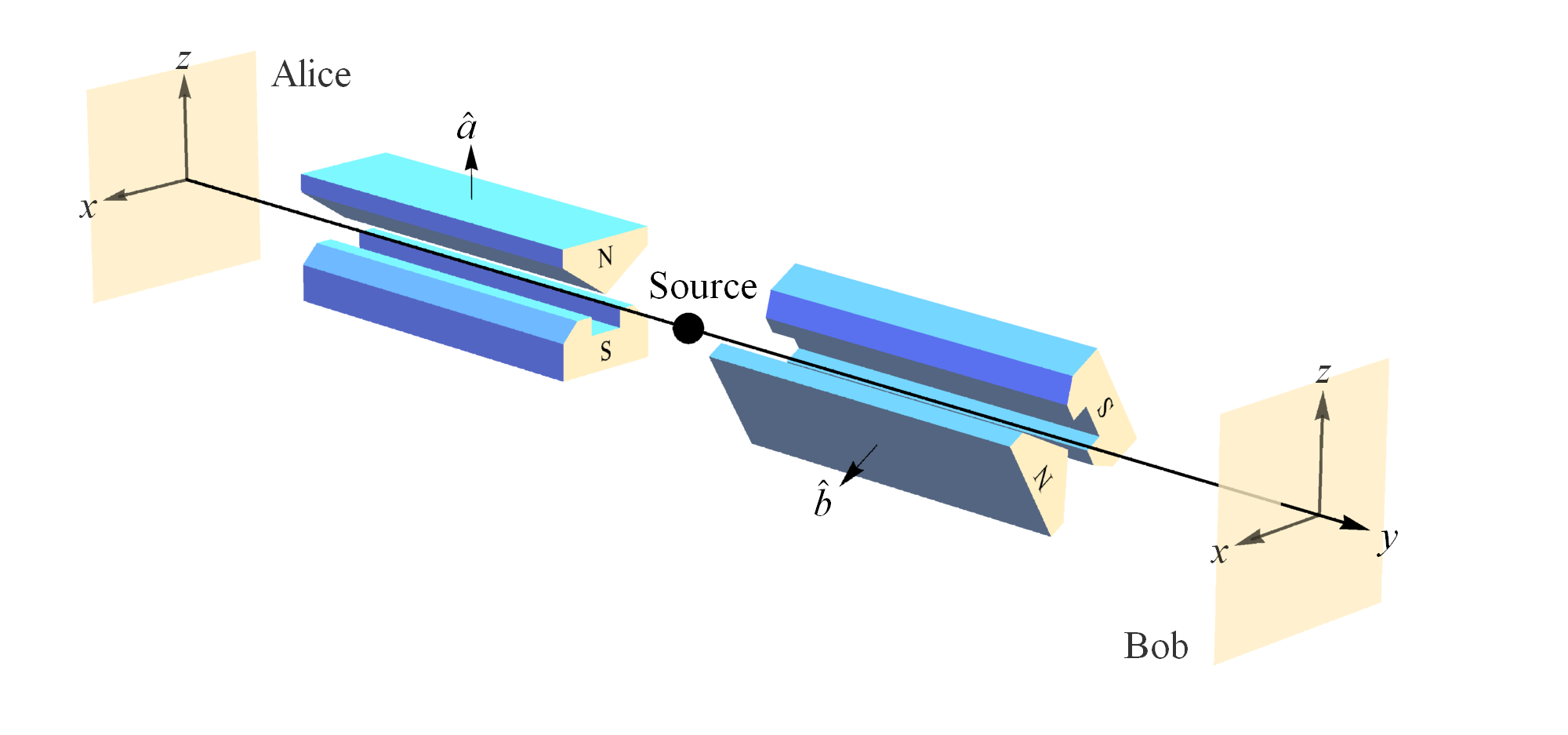}
\caption{Alice and Bob making spin measurements on a pair of spin-entangled particles with their Stern-Gerlach (SG) magnets and detectors in the $xz$-plane. Here Alice and Bob's SG magnets are not aligned so these measurements represent different reference frames. Since their outcomes satisfy Dakic and Brukner's Axiom 1 in all reference frames and satisfy explicit conservation of spin angular momentum in the same reference frame, they can only satisfy conservation of spin angular momentum on \textit{average} in different reference frames. This ``average-only'' conservation corresponds to the ``elliptope constraint'' of Janas et al. \cite{janas2019}} \label{EPRBmeasure}
\end{center}
\end{figure}

There are those in quantum information theory who have called for a principle(s) of a similar nature for quantum mechanics. Chris Fuchs writes \cite[p. 285]{fuchs1}:
\begin{quote}
Compare [quantum mechanics] to one of our other great physical theories, special relativity. One could make the statement of it in terms of some very crisp and clear physical principles: The speed of light is constant in all inertial frames, and the laws of physics are the same in all inertial frames. And it struck me that if we couldn't take the structure of quantum theory and change it from this very overt mathematical speak -- something that didn't look to have much physical content at all, in a way that anyone could identify with some kind of physical principle -- if we couldn't turn that into something like this, then the debate would go on forever and ever. And it seemed like a worthwhile exercise to try to reduce the mathematical structure of quantum mechanics to some crisp physical statements. \end{quote}
Herein, we make progress on that front by extending NPRF to include the measurement of another fundamental constant of nature, Planck's constant \textit{h}. As Steven Weinberg points out, measuring an electron's spin via Stern-Gerlach (SG) magnets constitutes the measurement of ``a universal constant of nature, Planck's constant'' \cite[p. 3]{weinberg2017} (Figure \ref{SGExp}). So if NPRF applies equally here, everyone must measure the same value for Planck's constant \textit{h} regardless of their SG magnet orientations relative to the source, which like the light postulate is an empirical fact. By ``relative to the source'' of a pair of spin-entangled particles, we mean relative ``to the vertical in the plane perpendicular to the line of flight of the particles'' \cite[p. 943]{mermin1981} (Figure \ref{EPRBmeasure}). Here the possible spin outcomes $\pm\frac{\hslash}{2}$ represent a fundamental (indivisible) unit of information per Dakic and Brukner's first axiom in their reconstruction of quantum theory, ``An elementary system has the information carrying capacity of at most one bit'' \cite{dakic}. Thus, different SG magnet orientations relative to the source constitute different ``reference frames'' in quantum mechanics just as different velocities relative to the source constitute different ``reference frames'' in special relativity. Since NPRF leads to the counterintuitive aspects (``mysteries'') of time dilation and length contraction in special relativity, it is perhaps not surprising that NPRF produces a ``mystery'' for quantum mechanics associated with the measurement of \textit{h} as well.

As David Mermin pointed out \cite[p. 1]{mermin2019}: 
\begin{quote}
Everybody who has learned quantum mechanics agrees how to use it. `Shut up and calculate!' There is no ambiguity, no confusion, and spectacular success. What we lack is any consensus about what one is actually talking about as one uses quantum mechanics. There is an unprecedented gap between the abstract terms in which the theory is couched and the phenomena the theory enables us so well to account for. We do not understand the meaning of this strange conceptual apparatus that each of us uses so effectively to deal with our world.
\end{quote}
And Weinberg writes \cite[p. 2]{weinberg2017}:
\begin{quote}
Many physicists came to think that the reaction of Einstein and Feynman and others to the unfamiliar aspects of quantum mechanics had been overblown. This used to be my view. ... Even so, I’m not as sure as I once was about the future of quantum mechanics. It is a bad sign that those physicists today who are most comfortable with quantum mechanics do not agree with one another about what it all means.
\end{quote}
To which Mermin responds \cite[p. 12]{mermin2019}, ``Steven Weinberg shares my concern that the lack of agreement about the meaning of quantum mechanics is a warning that ought to be taken seriously.'' One of the reasons quantum mechanics is so strange is its prediction and verification of quantum entanglement.  

\begin{figure}[h!]
\begin{center}
\includegraphics [height = 50mm]{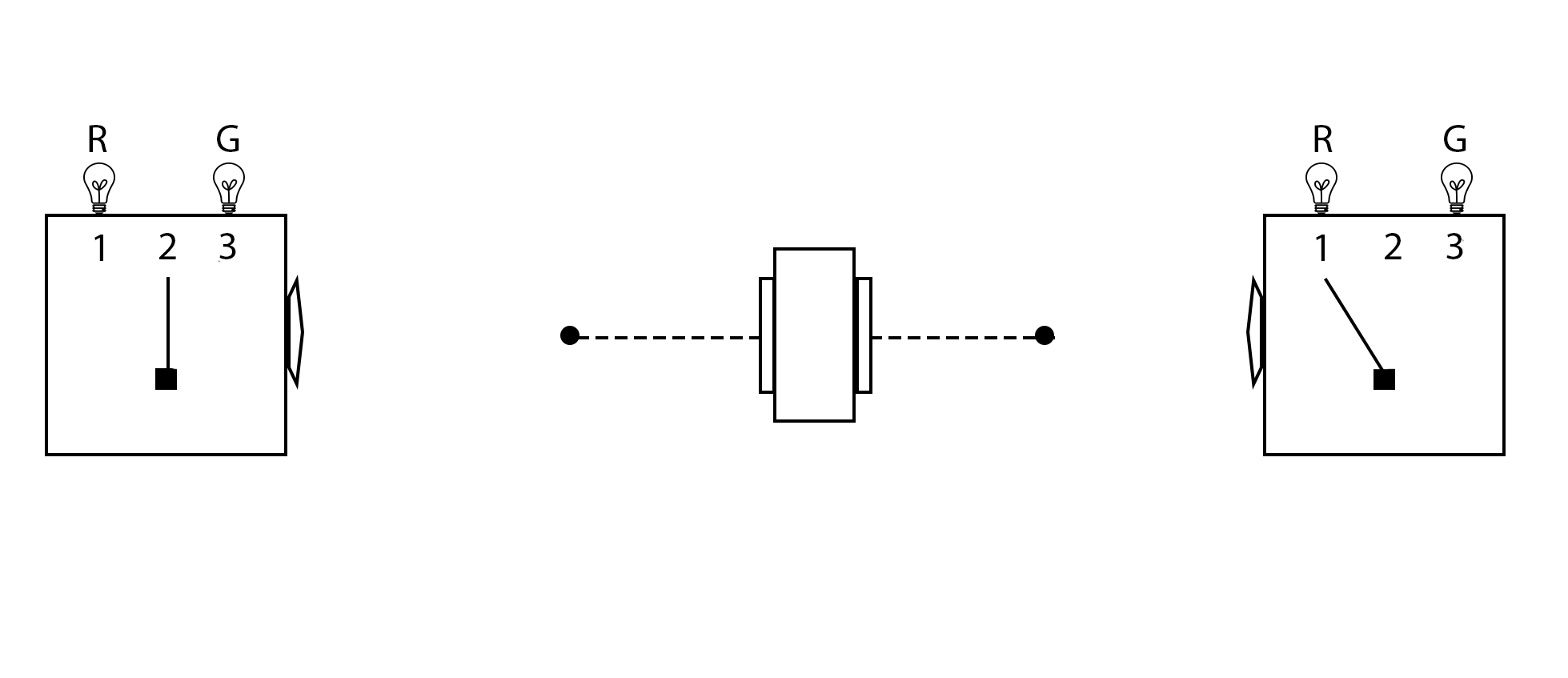}  \caption{\textbf{The Mermin Device}. Alice has her measuring device on the left set to 2 and Bob has his measuring device on the right set to 1. The particles have been emitted by the source in the middle and are in route to the measuring devices.} \label{mermin}
\end{center}
\end{figure}

In 1981, Mermin revealed the conundrum of quantum entanglement for a general audience \cite{mermin1981} using his ``simple device,'' which we will refer to as the ``Mermin device'' (Figure \ref{mermin}). Concerning this paper Richard Feynman wrote to Mermin, ``One of the most beautiful papers in physics that I know of is yours in the American Journal of Physics’’ \cite[p. 366-7]{feynman}. The Mermin device functions according to two facts concerning measurement outcomes in the same reference frame (``case (a)'') and measurement outcomes in different reference frames (``case (b)'') that are seemingly contradictory, thus the ``mystery.'' Mermin simply supplies these facts and shows the contradiction, which the ``general reader'' can easily understand. In other words, to understand the conundrum of the device required no knowledge of physics, just some simple probability theory, which made the presentation all the more remarkable. In subsequent publications, he ``revisited'' \cite{merminRevisited} and ``refined'' \cite{merminRefined} the ``mystery'' of quantum entanglement with similarly simple devices. In this paper, we will focus on the original Mermin device as it relates to the ``mystery'' of entanglement via the Bell spin states (Mermin's inspiration for his device), since it is particularly amenable to our resolution of the ``mystery'' that then provides a connection to special relativity via NPRF. 

Concerning his device Mermin wrote, ``Although this device has not been built, there is no reason in principle why it could not be, and probably no insurmountable practical difficulties'' \cite[p. 941]{mermin1981}. Sure enough, the experimental confirmation of the ``mystery'' of quantum entanglement is so common that it can now be carried out in the undergraduate physics laboratory \cite{dehlinger}. Thus, there is no disputing that the conundrum of the Mermin device has been experimentally well verified, vindicating its prediction by quantum mechanics.  

While the conundrum of the Mermin device is now a well-established fact, Mermin's challenge to ``translate the elementary quantum-mechanical reconciliation of cases (a) and (b) into terms meaningful to a general reader struggling with the dilemma raised by the device'' \cite[p. 943]{mermin1981} arguably remains unanswered. Of course, what Mermin desires is ``a picture of the way nature works'' or a ``crisp physical statement'' that is compelling and relatively easy to grasp. To answer this challenge, it is generally acknowledged that one needs a compelling model of physical reality or a compelling physical principle by which the conundrum of the Mermin device is resolved. Such a model needs to do more than the ``Copenhagen interpretation'' \cite{becker}, which Mermin characterized as ``shut up and calculate'' \cite{merminShutUp}. Concerning this ``shut up and calculate'' or ``instrumentalist'' approach to quantum mechanics, Weinberg writes \cite[p. 4]{weinberg2017}:
\begin{quote}
It seems to me that the trouble with this approach is not only that it gives up on an ancient aim of science: to say what is really going on out there. It is a surrender of a particularly unfortunate kind. In the instrumentalist approach, we have to assume, as fundamental laws of nature, the rules (such as the Born rule I mentioned earlier) for using the wave function to calculate the probabilities of various results when humans make measurements. Thus humans are brought into the laws of nature at the most fundamental level. 
\end{quote}
In other words, while the ``elementary quantum-mechanical reconciliation of cases (a) and (b)'' accurately predicts the conundrum, the formalism itself does not provide a model of physical reality or underlying physical principle to resolve the conundrum, compelling or otherwise. Thus, a satisfactory answer to Mermin's challenge will certainly help us ``say what is really going on out there.''

Janas et al. \cite{janas2019} recently supplied the ``elliptope constraint'' for the Mermin device using correlation arrays \textit{a la} Jeff Bub's book \textit{Bananaworld} \cite{bubbook}. This constraint allows for a geometrical representation of ``the class of correlations allowed by quantum mechanics in this setup as an elliptope in a non-signaling cube'' \cite[p. 1]{janas2019}. They then use ``raffles with baskets of tickets'' to find the subspace of the quantum elliptope occupied by local hidden-variable theories. They found that such correlations ``can be represented geometrically by a tetrahedron contained within the elliptope'' \cite[p. 1]{janas2019}. Raffles, monkeys, and bananas are conceptually accessible to the ``general reader'' and the resulting nested geometrical figures (tetrahedron for classical correlations inside elliptope for quantum correlations inside a non-signaling cube) provides a nice visualization of the ``mystery'' of the Mermin device. The Janas et al. interpretation of quantum mechanics is based on ``probabilities and expectation values ... determined by inner products of vectors in Hilbert space'' \cite[p. 1]{janas2019}. Herein, we will make their elliptope constraint a bit more accessible by revealing a counterpart to it in real space that we call ``average-only'' conservation. 

As we will show, this ``average-only'' conservation is ``conservation per NPRF.'' Thus, NPRF provides a deeper understanding of ``average-only'' conservation and the elliptope constraint, and directly relates the ``mysteries'' of time dilation and length contraction in special relativity to the ``mystery'' of Bell spin state entanglement in quantum mechanics per the restricted Lorentz symmetry group. We will also show how this answer to Mermin's challenge complements his current view of the meaning of quantum mechanics per QBism and how it answers Weinberg's question, ``how do probabilities get into quantum mechanics?'' Note, this answer to Mermin's challenge does not mean ``humans are brought into the laws of nature at the most fundamental level,'' as different SG magnet orientations relative to the source and different velocities relative to the source do not imply the necessity of human observation. Additionally, the principle of NPRF reveals an underlying coherence between non-relativistic quantum mechanics and special relativity where others have perceived tension \cite{bellbook,mamone}. For all these reasons, we believe ``conservation per NPRF'' is a ``crisp physical statement'' that contributes to ``building a picture of the way nature works'' in order to ``say what is really going on out there,'' thus providing progress on a desideratum of quantum information theorists.

\section*{The Mermin Device and Its Conundrum} \label{Secmermindevice}
Here we remind the reader how the Mermin device works and how it relates to the spin measurements carried out with SG magnets and detectors (Figures \ref{SGExp} \& \ref{EPRBmeasure}). The exposition of the ``mystery'' and our resolution thereof are accessible to the ``general reader'' who has taken a first course in physics. In Methods, we provide technical details for the interested reader. 

\begin{figure}[h!]
\begin{center}
\includegraphics [height = 40mm]{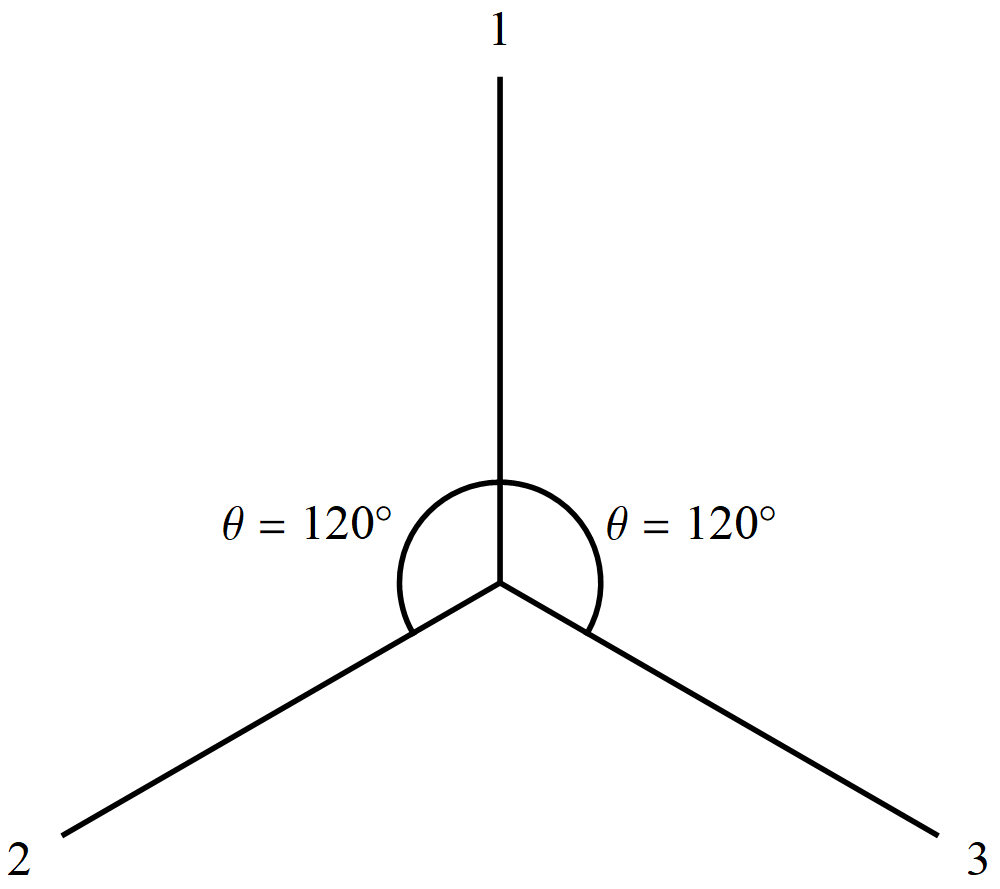} 
\caption{Three possible orientations of Alice and Bob's SG magnets for the Mermin device.} \label{SGorientations}
\end{center}
\end{figure}

The Mermin device contains a source (middle box in Figure \ref{mermin}) that emits a pair of spin-entangled particles towards two detectors (boxes on the left and right in Figure \ref{mermin}) in each trial of the experiment. We will focus formally on spin-$\frac{1}{2}$ particles herein, but his device is also valid conceptually for spin-1 particles \cite{janas2019,TsirelsonBound2019}. The settings (1, 2, or 3) on the left and right detectors are controlled randomly by Alice and Bob, respectively, and each measurement at each detector produces either a result of R or G. The following two facts obtain (Table \ref{tb:1}):
\begin{enumerate}
    \item When Alice and Bob's settings are the same in a given trial (``case (a)''), their outcomes are always the same, $\frac{1}{2}$ of the time RR (Alice's outcome is R and Bob's outcome is R) and $\frac{1}{2}$ of the time GG (Alice's outcome is G and Bob's outcome is G).
    \item When Alice and Bob's settings are different in a given trial (``case (b)''), the outcomes are the same $\frac{1}{4}$ of the time, $\frac{1}{8}$ RR and $\frac{1}{8}$ GG. 
\end{enumerate}
The two possible Mermin device outcomes R and G represent two possible spin measurement outcomes ``up'' and ``down,'' respectively, (Figure \ref{SGExp}) and the three possible Mermin device settings represent three different orientations of the SG magnets (Figures \ref{EPRBmeasure} \& \ref{SGorientations}). Mermin writes \cite[p. 942]{mermin1981}: 
\begin{quote}
Why do the detectors always flash the same colors when the switches are in the same positions? Since the two detectors are unconnected there is no way for one to ``know'' that the switch on the other is set in the same position as its own.
\end{quote}
This leads him to introduce ``instruction sets'' to account for the behavior of the device when the detectors have the same settings. Concerning the use of instruction sets to account for Fact 1 he writes, ``It cannot be proved that there is no other way, but I challenge the reader to suggest any'' \cite[p. 942]{mermin1981}. Mermin explicitly excludes the possibilities of retrocausality and superluminal communication between the particles. That is, the particles cannot ``know'' what settings they will encounter until they arrive at the detectors and they cannot communicate their settings and outcomes with each other in spacelike fashion. Now look at all trials when Alice's particle has instruction set RRG and Bob's has instruction set RRG, for example. 

\begin{table}
\begin{center}
\begin{tabular}{ccc}
\textbf{Case (a) Same Settings} && \textbf{Case (b) Different Settings}\\
    \begin{tabular}{cc|cc}
    \multicolumn{2}{c}{}&\multicolumn{2}{c}{Alice} \\
    && R & G \\
    \cline{2-4}
    \multirow{2}*{Bob} & R & 1/2 & 0 \\
    & G &  0 & 1/2 
    \end{tabular}
& \hspace*{0.5in} &
    \begin{tabular}{cc|cc}
    \multicolumn{2}{c}{}&\multicolumn{2}{c}{Alice} \\
    && R & G \\
    \cline{2-4}
    \multirow{2}*{Bob} & R & 1/8 & 3/8 \\
    & G &  3/8 & 1/8 \\
    \end{tabular} \\
\end{tabular}
\caption{\textbf{Summary of outcome probabilities for the Mermin device}. These are in accord with the Malus law.}
\label{tb:1}
\end{center}
\end{table}

That means Alice and Bob's outcomes in setting 1 will both be R, in setting 2 they will both be R, and in setting 3 they will both be G. That is, the particles will produce an RR result when Alice and Bob both choose setting 1 (referred to as ``11''), an RR result when both choose setting 2 (referred to as ``22''), and a GG result when both choose setting 3 (referred to as ``33''). That is how instruction sets guarantee Fact 1. For different settings Alice and Bob will obtain the same outcomes when Alice chooses setting 1 and Bob chooses setting 2 (referred to as ``12''), which gives an RR outcome. And, they will obtain the same outcomes when Alice chooses setting 2 and Bob chooses setting 1 (referred to as ``21''), which also gives an RR outcome. That means we have the same outcomes for different settings in 2 of the 6 possible case (b) situations, i.e., in $\frac{1}{3}$ of case (b) trials for this instruction set. This $\frac{1}{3}$ ratio holds for any instruction set with two R(G) and one G(R). 

\begin{table}
\begin{center}
\begin{tabular}{ccc}
\textbf{Case (a) Same Settings} && \textbf{Case (b) Different Settings}\\
\begin{tabular}{cc|cc}
    \multicolumn{2}{c}{}&\multicolumn{2}{c}{Alice} \\
    && R & G \\
    \cline{2-4}
    \multirow{2}*{Bob} & R & 1/2 & 0 \\
    & G &  0 & 1/2 
    \end{tabular}
& \hspace*{0.5in} &
    \begin{tabular}{cc|cc}
    \multicolumn{2}{c}{}&\multicolumn{2}{c}{Alice} \\
    && R & G \\
    \cline{2-4}
    \multirow{2}*{Bob} & R & 1/4 & 1/4 \\
    & G &  1/4 & 1/4 \\
    \end{tabular} \\
\end{tabular}
\caption{\textbf{Summary of outcome probabilities for instruction sets}. We are assuming the eight possible instruction sets are produced with equal frequency.}
\label{tb:2}
\end{center}
\end{table}

The only other possible instruction sets are RRR or GGG where Alice and Bob's outcomes will agree in $\frac{9}{9}$ of all trials. Thus, the ``Bell inequality'' \cite{bell} for the Mermin device says that instruction sets must produce the same outcomes in more than $\frac{1}{3}$ of all case (b) trials. Indeed, if all eight instruction sets are produced with equal frequency, the RR, GG, RG, and GR outcomes for any given pair of unlike settings (12, 13, 21, 23, 31, or 32) will be produced in equal numbers, so the probability of getting the same outcomes for different settings is $\frac{1}{2}$ (Table \ref{tb:2}). But, Fact 2 for quantum mechanics says you only get the same outcomes in $\frac{1}{4}$ of all those trials, thereby violating the prediction per instruction sets. Thus, the conundrum of Mermin's device is that the instruction sets needed for Fact 1 fail to yield the proper outcomes for Fact 2. 

That quantum mechanics accurately predicts the observed phenomenon without spelling out any means \textit{a la} instruction sets for how it works prompted Lee Smolin to write \cite[p. xvii]{smolin}: 
\begin{quote}
I hope to convince you that the conceptual problems and raging disagreements that have bedeviled quantum mechanics since its inception are unsolved and unsolvable, for the simple reason that the theory is wrong. It is highly successful, but incomplete. 
\end{quote}
Of course, this is precisely the complaint leveled by Einstein, Podolsky, and Rosen in their famous paper, ``Can Quantum-Mechanical Description of Physical Reality Be Considered Complete?'' \cite{EPRpaper}. Our point herein is that quantum entanglement does not render quantum mechanics wrong or incomplete. There is no disputing that quantum mechanics is a tremendously successful theory and as we will show, it is as complete as possible given that it must conform to NPRF.

So, Mermin's challenge to the ``physicist reader'' is to explain to the ``general reader'' how quantum mechanics reconciles Facts 1 and 2. We will answer Mermin's challenge by showing that Facts 1 and 2 follow from a very reasonable conservation principle and thereby render Smolin's sentiment entirely misguided. That is, we will see that quantum mechanics is not only complete, but it shares an underlying coherence with Einstein's other revolution \cite{smolin}, special relativity, i.e., the ``mysteries'' of both are grounded in the same principle, ``no preferred reference frame.'' The reasonable conservation principle resides in the correlation function, so we start there.

The correlation function between two outcomes over many trials is the average of the two values multiplied together. In this case, there are only two possible outcomes for any setting, +1 (up or R) or --1 (down or G), so the largest average possible is +1 (total correlation, RR or GG, as when the settings are the same) and the smallest average possible is --1 (total anti-correlation, RG or GR). One way to write the equation for the correlation function is
\begin{equation}
\langle \alpha,\beta \rangle = \sum (i \cdot j) \cdot p(i,j \mid \alpha,\beta)  \label{average}
\end{equation}
where $p(i,j \mid \alpha,\beta)$ is the probability that Alice measures $i$ and Bob measures $j$ given that Alice's SG magnets are at angle $\alpha$ and Bob's SG magnets are at angle $\beta$, and $(i \cdot j)$ is just the product of the outcomes $i$ and $j$. The correlation function for instruction sets for case (a) is the same as that of the Mermin device for case (a), i.e., they're both 1. Thus, we must explore the difference between the correlation function for instruction sets and the Mermin device for case (b).

To get the correlation function for instruction sets for case (b), we need the probabilities of measuring the same outcomes and different outcomes for different settings, so we can use Eq. (\ref{average}). We saw that when we had two R(G) and one G(R), the probability of getting the same outcomes for different settings was $\frac{1}{3}$ (this would break down to $\frac{1}{6}$ for each of RR and GG overall). Thus, the probability of getting different outcomes would be $\frac{2}{3}$ for these types of instruction sets ($\frac{1}{3}$ for each of RG and GR). That gives a correlation function of 
\begin{multline}
\langle \alpha,\beta \rangle = \left(+1\right)\left(+1\right)\left(\frac{1}{6}\right) + \left(-1\right)\left(-1\right)\left(\frac{1}{6}\right) + \left(+1\right)\left(-1\right)\left(\frac{2}{6}\right) +  \left(-1\right)\left(+1\right)\left(\frac{2}{6}\right)= -\frac{1}{3}  
\end{multline} 
For the other type of instruction sets, RRR and GGG, we would have a correlation function of +1 for different settings, so overall the correlation function for instruction sets for case (b) has to be larger than $-\frac{1}{3}$. Again, if all eight instruction sets are produced with equal frequency, the probability for any particular outcome is $\frac{1}{4}$ for case (b) (Table \ref{tb:2}) giving a correlation function of zero. That means the results are uncorrelated as one would expect given that all possible instruction sets are produced randomly. From this we would typically infer that there is nothing that needs to be explained. Indeed, if Fact 1 about case (a) obtains due to some underlying conservation principle at the source, then uncorrelated results for case (b) is more surprising than the anti-correlated results that we now show obtain per the Mermin device. In other words, instruction sets entail there are no observable case (b) consequences for the case (a) conservation. As we now show, the Mermin device says otherwise.

Fact 2 for the Mermin device says the probability of getting the same results (RR or GG) for different settings is $\frac{1}{4}$ ($\frac{1}{8}$ for each of RR and GG, Table \ref{tb:1}). Thus, the probability of getting different outcomes for different settings must be $\frac{3}{4}$ ($\frac{3}{8}$ for each of RG and GR, Table \ref{tb:1}). That gives a correlation function of 
\begin{multline}
\langle \alpha,\beta \rangle = \left(+1\right)\left(+1\right)\left(\frac{1}{8}\right) + \left(-1\right)\left(-1\right)\left(\frac{1}{8}\right) + \left(+1\right)\left(-1\right)\left(\frac{3}{8}\right) + \left(-1\right)\left(+1\right)\left(\frac{3}{8}\right)= -\frac{1}{2}  \label{MerminCorr}
\end{multline} 
That means the Mermin device is more strongly anti-correlated for different settings than instruction sets. Indeed, again, if all possible instruction sets are produced with equal frequency, the Mermin device evidences something to explain (anti-correlated results for case (b)) where instruction sets suggest there is nothing in need of explanation (uncorrelated results for case (b)). Again, the Mermin device indicates that the conservation principle responsible for Fact 1 of case (a) has observable implications (Fact 2) for case (b) while instruction sets say we should not expect to see any consequence of Fact 1 for case (b). Mermin's challenge then amounts to providing a compelling physical model or compelling physical principle to account for Facts 1 and 2 for case (a) and case (b), respectively.

\section*{The Bell Spin States}\label{Secbellstates}
In order to ``translate the elementary quantum-mechanical reconciliation of cases (a) and (b),'' we first provide an accessible introduction to that ``elementary quantum-mechanical reconciliation'' for the ``general reader.'' [The technical details are provided in Methods for the interested ``physicist reader.''] This amounts to a review of the nature of conservation at work in the Bell spin states for spin-$\frac{1}{2}$ particles as revealed by the correlation function. Essentially, there are four combinations of conserved spin angular momentum represented by the four Bell spin states for the pair of spin-entangled particles
\begin{equation}
\begin{aligned}
&|\psi_-\rangle = \frac{|ud\rangle \,- |du\rangle}{\sqrt{2}}\\
&|\psi_+\rangle = \frac{|ud\rangle + |du\rangle}{\sqrt{2}}\\
&|\phi_-\rangle = \frac{|uu\rangle \,- |dd\rangle}{\sqrt{2}}\\
&|\phi_+\rangle = \frac{|uu\rangle + |dd\rangle}{\sqrt{2}}\\ \label{BellStates}
\end{aligned}
\end{equation}
where $u$ represents an up outcome and $d$ represents a down outcome for the SG measurements (Figures \ref{SGExp} \& \ref{EPRBmeasure}). 

The first state $|\psi_-\rangle$ is called the ``spin singlet state'' and it represents a conserved spin angular momentum of zero ($S = 0$, particles' spin angular momenta are anti-aligned) for the two particles involved. Specifically, $|\psi_-\rangle$ says that when the SG magnets are aligned (Alice and Bob are in the same reference frame) the outcomes are always opposite ($\frac{1}{2}$ $ud$ and $\frac{1}{2}$ $du$). This conservation holds as Alice and Bob rotate their SG magnets together in any plane of real space, i.e., the conserved $S = 0$ state is rotationally (SO(3)) invariant in any plane of real space. 

The other three states are called the ``spin triplet states'' and they each represent a conserved, rotationally invariant spin angular momentum of one ($S = 1$ in units of $\hslash = 1$, particles' spin angular momenta are aligned) in a particular plane of real space. Specifically, $|\phi_+\rangle$ is in the $xz$-plane, $|\phi_-\rangle$ is in the $yz$-plane, and $|\psi_+\rangle$ is in the $xy$-plane of real space (again, details are in Methods for the interested reader). So, when the SG magnets are aligned (the measurements are being made in the same reference frame) anywhere in the respective plane of symmetry the outcomes are always the same ($\frac{1}{2}$ $uu$ and $\frac{1}{2}$ $dd$). It is a planar conservation and our experiment would determine which plane, e.g., ``the plane perpendicular to the line of flight of the particles'' for the Mermin device. If you want to model a conserved $S = 1$ for some other plane, you simply create a superposition, i.e., expand in the spin triplet basis.  In all four cases, the entanglement represents the conservation of spin angular momentum for the process creating the state. Now let us relate this to the correlation functions. 

The Pauli spin matrices are used for the spin measurement operators $\sigma_x$, $\sigma_y$, and $\sigma_z$, so that if Alice is making her spin measurement $\sigma_1$ in the $\hat{a}$ direction and Bob is making his spin measurement $\sigma_2$ in the $\hat{b}$ direction (Figure \ref{EPRBmeasure}), we have
\begin{equation}
\begin{aligned}
    &\sigma_1 = \hat{a}\cdot\vec{\sigma}=a_x\sigma_x + a_y\sigma_y + a_z\sigma_z \\
    &\sigma_2 = \hat{b}\cdot\vec{\sigma}=b_x\sigma_x + b_y\sigma_y + b_z\sigma_z \\ \label{sigmas}
\end{aligned}
\end{equation}
Using this formalism and the fact that $\{|uu\rangle,|ud\rangle,|du\rangle,|dd\rangle\}$ is an orthonormal set ($\langle uu|uu\rangle = 1$, $\langle uu|ud\rangle = 0$, $\langle du|du\rangle = 1$, etc.), you can show that the correlation functions are given by
\begin{equation}
\begin{aligned}
&\langle\psi_-|\sigma_1\sigma_2|\psi_-\rangle = &-a_xb_x - a_yb_y - a_zb_z\\
&\langle\psi_+|\sigma_1\sigma_2|\psi_+\rangle = &a_xb_x + a_yb_y - a_zb_z\\
&\langle\phi_-|\sigma_1\sigma_2|\phi_-\rangle = &-a_xb_x + a_yb_y + a_zb_z\\
&\langle\phi_+|\sigma_1\sigma_2|\phi_+\rangle = &a_xb_x - a_yb_y + a_zb_z\\ \label{gencorrelations}
\end{aligned}
\end{equation}
That is to say, the correlation function for the spin singlet state is $\langle\psi_-|\sigma_1\sigma_2|\psi_-\rangle = -\cos{(\theta)}$ where $\theta$ is the angle between $\hat{a}$ and $\hat{b}$. The correlation functions for the spin triplet states are $\langle\psi_+|\sigma_1\sigma_2|\psi_+\rangle = \cos{(\theta})$ where $\theta$ is the angle between $\hat{a}$ and $\hat{b}$ in the $xy$-plane of symmetry, $\langle\phi_-|\sigma_1\sigma_2|\phi_-\rangle = \cos{(\theta})$ where $\theta$ is the angle between $\hat{a}$ and $\hat{b}$ in the $yz$-plane of symmetry, and $\langle\phi_+|\sigma_1\sigma_2|\phi_+\rangle = \cos{(\theta)}$ where $\theta$ is the angle between $\hat{a}$ and $\hat{b}$ in the $xz$-plane of symmetry. 

There is a simple analogy here with special relativity. When Alice and Bob have different velocities relative to the source (occupy different reference frames), the corresponding Lorentz transformations depend only on their relative velocity. Here, when Alice and Bob have different SG magnet orientations relative to the source (occupy different reference frames), the resulting correlation functions depend only on their relative SG orientation angle.

It is important to note that the conservation at work here deals with the measurement outcomes proper. Per Dakic and Brukner's axiomatic reconstruction of quantum theory \cite{dakic}, the Bell spin states represent measurement outcomes on an entangled pair of ``elementary systems.'' Axiom 1 of their reconstruction states, ``An elementary system has the information carrying capacity of at most one bit.'' Thus, it is not the case that the measurement outcomes are merely the revealed portion of a greater wealth of information carried by an underlying quantum system. Colloquially put, Alice and Bob's measurement outcomes exhaust the available information, there is nothing ``hidden.''

In conclusion, the correlation function for any pair of case (b) settings in the Mermin device (Figures \ref{mermin} \& \ref{SGorientations}) is $\cos(120^\circ) = -\frac{1}{2}$, in agreement with Eq. (\ref{MerminCorr}), instead of zero per that of instruction sets. In other words, the Mermin device represents spin measurements on an $S = 1$ spin-entangled pair of particles in their plane of symmetry in real space at the angles given by Figure \ref{SGorientations}. If you let Bob's R(G) results represent Alice's G(R) results, the Mermin device then represents spin measurements on an $S = 0$ spin-entangled pair of particles in some plane of real space (all planes are planes of symmetry for $S = 0$). In that case, the correlation function for any pair of case (b) settings in the Mermin device is $-\cos(120^\circ) = \frac{1}{2}$, instead of zero per that of instruction sets. So, for the $S = 0$ case (b) situation, the Mermin device is giving us correlated results rather than uncorrelated results per instruction sets. And, for the $S = 1$ case (b) situation, the Mermin device is giving us anti-correlated results rather than uncorrelated results per instruction sets. We now ``translate [this] elementary quantum-mechanical reconciliation of cases (a) and (b) into terms meaningful to a general reader'' and thereby ``say what is really going on out there.''

\section*{Average-Only Conservation} \label{Secprinciple}
Now that we understand the ``elementary quantum-mechanical reconciliation of cases (a) and (b),'' it turns out that the ``physicist reader'' can ``translate'' it ``into terms meaningful to a general reader'' rather easily. This explanation is accessible to any ``general reader'' who understands the conservation of angular momentum. Let us start with the quantum correlation function for the spin singlet state \cite{unnikrishnan}. 

Again, the total spin angular momentum is zero and every measurement produces outcomes of $+1$ (up) or $-1$ (down) in units of $\frac{\hslash}{2} = 1$. Alice and Bob both measure $+1$ and $-1$ results with equal frequency for any SG magnet angle and when their angles are equal (case (a)) they obtain different outcomes giving total spin angular momentum of zero. This result is not difficult to understand via conservation of spin angular momentum, because Alice and Bob's measured values of spin angular momentum cancel directly when $\alpha = \beta$ (Figure \ref{EPRBmeasure}). But, when Bob's SG magnets are rotated by $\alpha - \beta = \theta$ relative to Alice's SG magnets (case (b)), we need to clarify the situation.

We have two sets of data, Alice's set and Bob's set. They were collected in $N$ pairs (data events) with Bob's(Alice's) SG magnets at $\theta$ relative to Alice's(Bob's). We want to compute the correlation function for these $N$ data events which is

\begin{equation}\langle \alpha,\beta \rangle =\frac{(+1)_A(-1)_B + (+1)_A(+1)_B + (-1)_A(-1)_B + ...}{N}\end{equation}
Now partition the numerator into two equal subsets per Alice's equivalence relation, i.e., Alice's $+1$ results and Alice's $-1$ results

\begin{equation}\langle \alpha,\beta \rangle =\frac{(+1)_A(\sum \mbox{BA+})+(-1)_A(\sum \mbox{BA-})}{N}\end{equation}
where $\sum \mbox{BA+}$ is the sum of all of Bob's results (event labels) corresponding to Alice's $+1$ result (event label) and $\sum \mbox{BA-}$ is the sum of all of Bob's results (event labels) corresponding to Alice's $-1$ result (event label). Notice this is all independent of the formalism of quantum mechanics. Now, we rewrite that equation as
\begin{equation}\langle \alpha,\beta \rangle = \frac{1}{2}(+1)_A\overline{BA+} + \frac{1}{2}(-1)_A\overline{BA-} \label{consCorrel}\end{equation}
with the overline denoting average. Again, this correlation function is independent of the formalism of quantum mechanics. All we have assumed is that Alice and Bob measure $+1$ or $-1$ with equal frequency at any setting in computing this correlation function. Notice that to understand the quantum correlation responsible for Fact 2 of the Mermin device, i.e., the Fact that represents the deviation between the quantum and the classical correlations, we need to understand the origin of $\overline{BA+}$ and $\overline{BA-}$ for the Bell spin states. We now show what that is for the spin singlet state, then we extend the argument to the spin triplet states and underwrite it all with NPRF.

\begin{figure}[h]
\begin{center}
\includegraphics [height = 25mm]{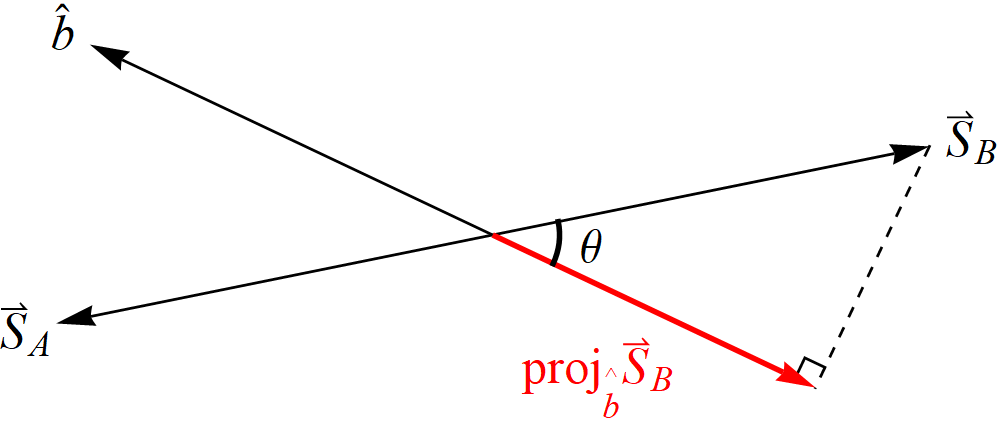}
\caption{The angular momentum of Bob's particle $\vec{S}_B = -\vec{S}_A$ projected along his measurement direction $\hat{b}$. This does \textit{not} happen with spin angular momentum.} \label{Projection1}
\end{center}
\end{figure}

\begin{figure}[h]
\begin{center}
\includegraphics [height = 25mm]{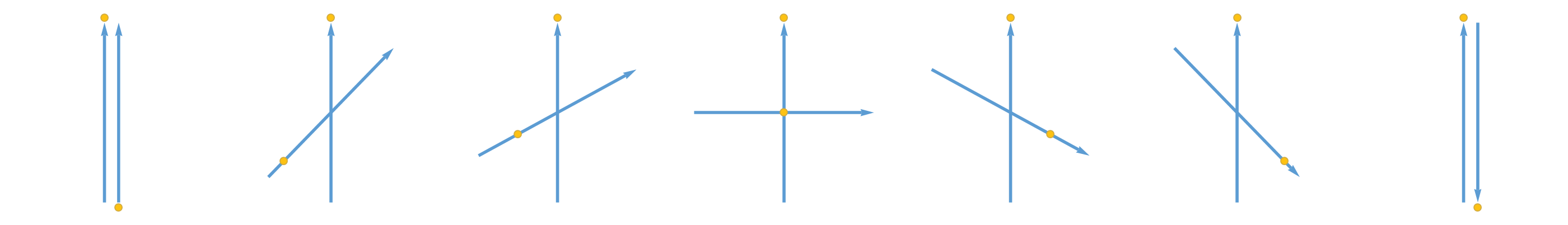}  
\caption{\textbf{Average View for the Spin Singlet State}. Reading from left to right, as Bob rotates his SG magnets relative to Alice's SG magnets for her $+1$ outcome, the average value of his outcome varies from $-1$ (totally down, arrow bottom) to 0 to +1 (totally up, arrow tip). This obtains per conservation of spin angular momentum on average in accord with no preferred reference frame. Bob can say exactly the same about Alice's outcomes as she rotates her SG magnets relative to his SG magnets for his $+1$ outcome. That is, their outcomes can only satisfy conservation of spin angular momentum on average in different reference frames, because they only measure $\pm 1$, never a fractional result. Thus, just as with the light postulate of special relativity, we see that no preferred reference frame leads to a counterintuitive result. Here it requires quantum outcomes $\pm 1 \left(\frac{\hslash}{2}\right)$ for all measurements and that leads to the ``mystery'' of ``average-only'' conservation. Note: Here you can see the physical reason that $\theta = 2\Theta$ for spin-$\frac{1}{2}$ particles found in Methods, i.e., spin is a bi-directional property in the plane of symmetry for spin-$\frac{1}{2}$ particles.} \label{AvgViewSinglet}
\end{center}
\end{figure}

In classical physics, one would say the projection of the spin angular momentum vector of Alice's particle $\vec{S}_A = +1\hat{a}$ along $\hat{b}$ is $\vec{S}_A\cdot\hat{b} = +\cos({\theta})$ where again $\theta$ is the angle between the unit vectors $\hat{a}$ and $\hat{b}$. That's because the prediction from classical physics is that all values between $+1 \left(\frac{\hslash}{2}\right)$ and $-1 \left(\frac{\hslash}{2}\right)$ are possible outcomes for a spin measurement (Figure \ref{SGExp}). From Alice's perspective, had Bob measured at the same angle, i.e., $\beta = \alpha$, he would have found the spin angular momentum vector of his particle was $\vec{S}_B = -\vec{S}_A = -1\hat{a}$, so that $\vec{S}_A + \vec{S}_B = \vec{S}_{Total} = 0$. Since he did not measure the spin angular momentum of his particle at the same angle, he should have obtained a fraction of the length of $\vec{S}_B$, i.e., $\vec{S}_B\cdot\hat{b} = -1\hat{a}\cdot\hat{b} = -\cos({\theta})$ (Figure \ref{Projection1}; this also follows from counterfactual spin measurements on the single-particle state \cite{boughn}). Of course, Bob only ever obtains $+1$ or $-1$, but suppose that Bob's outcomes \textit{average} $-\cos({\theta})$, which can certainly happen for a collection of $+1$ and $-1$ outcomes (Figure \ref{AvgViewSinglet}). This means 
\begin{equation}\overline{BA+} = -\cos(\theta) \label{AvgPlus1}\end{equation}
Likewise, for Alice's $(-1)_A$ results we have
\begin{equation}\overline{BA-} = \cos(\theta) \label{AvgMinus1}\end{equation}
Putting these into Eq. (\ref{consCorrel}) we obtain
\begin{equation}\langle \alpha,\beta \rangle = \frac{1}{2}(+1)_A(-\cos(\theta)) + \frac{1}{2}(-1)_A(\cos(\theta)) = -\cos(\theta) \label{correlEnd}\end{equation}
which is precisely the correlation function given by quantum mechanics for the spin singlet state as shown above. Notice that the ``average-only'' conservation of Eqs. (\ref{AvgPlus1} \& \ref{AvgMinus1}) is simply a mathematical fact for obtaining the quantum correlation function. Of course, Bob could partition the data according to his equivalence relation (per his reference frame) and claim that it is Alice who must average her results (obtained in her reference frame) to conserve spin angular momentum. Now for the spin triplet states.

\begin{figure}[h]
\begin{center}
\includegraphics [height = 35mm]{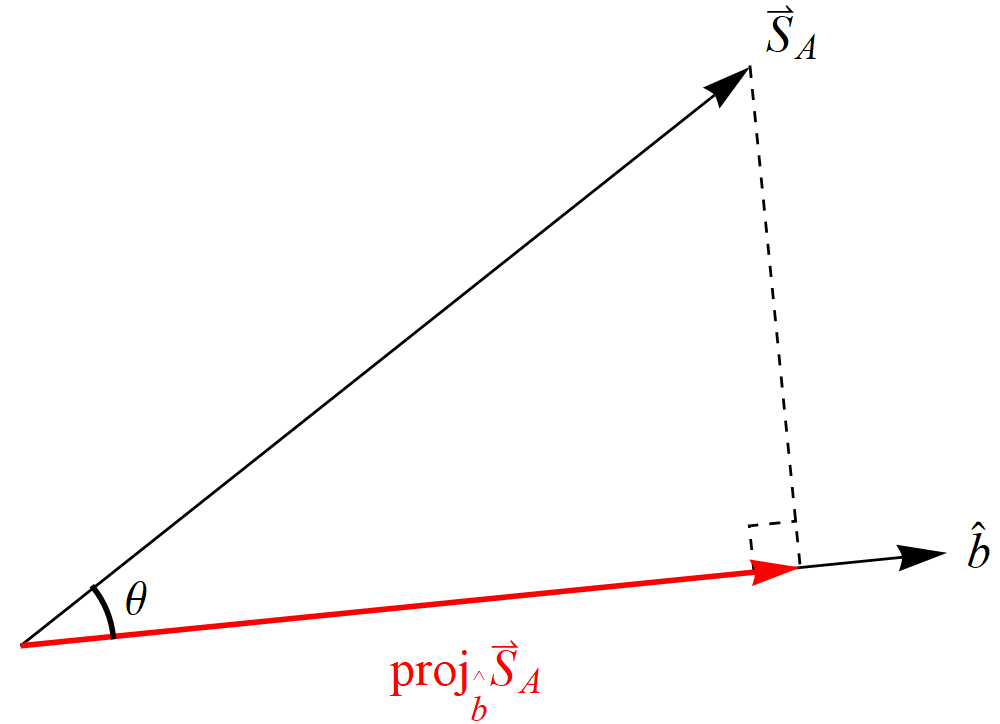} \caption{The angular momentum of Bob's particle $\vec{S}_B = \vec{S}_A$ projected along his measurement direction $\hat{b}$. This does \textit{not} happen with spin angular momentum.} \label{Projection}
\end{center}
\end{figure}

\begin{figure}[h]
\begin{center}
\includegraphics [height = 25mm]{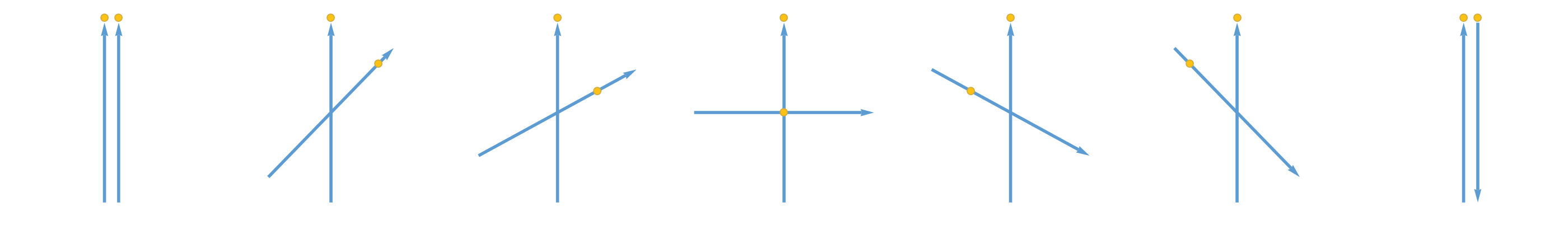}  \caption{\textbf{Average View for the Spin Triplet States}. Reading from left to right, as Bob rotates his SG magnets relative to Alice's SG magnets for her $+1$ outcome, the average value of his outcome varies from $+1$ (totally up, arrow tip) to $0$ to $-1$ (totally down, arrow bottom). This obtains per conservation of spin angular momentum on average in accord with no preferred reference frame. Bob can say exactly the same about Alice's outcomes as she rotates her SG magnets relative to his SG magnets for his $+1$ outcome. That is, their outcomes can only satisfy conservation of spin angular momentum on average in different reference frames, because they only measure $\pm 1$, never a fractional result. Again, just as with the light postulate of special relativity, we see that no preferred reference frame leads to a counterintuitive result. Here it requires quantum outcomes $\pm 1 \left(\frac{\hslash}{2}\right)$ for all measurements leading to the ``mystery'' of ``average-only'' conservation.} \label{AvgViewTriplet}
\end{center}
\end{figure}

As we saw above, the spin triplet states represent ``SO(3) conservation'' of spin angular momentum analogous to the spin singlet state. Thus, we can repeat our story for the $S = 1$ plane of SO(3) rotational invariance, whatever that is. From Alice's perspective, had Bob measured at the same angle, i.e., $\beta = \alpha$, he would have found the spin angular momentum vector of his particle was $\vec{S}_B = \vec{S}_A = +1\hat{a}$, so that $\vec{S}_A + \vec{S}_B = \vec{S}_{Total} = 2$ (this is S = 1 in units of $\frac{\hslash}{2} = 1$). Since he did not measure the spin angular momentum of his particle at the same angle, he should have obtained a fraction of the length of $\vec{S}_B$, i.e., $\vec{S}_B\cdot\hat{b} = +1\hat{a}\cdot\hat{b} = \cos{(\theta)}$ (Figure \ref{Projection}). Of course, Bob only ever obtains $+1$ or $-1$, but again suppose that Bob's outcomes \textit{average} $\cos{(\theta)}$ (Figures \ref{AvgViewTriplet} \& \ref{4Dpattern}). This means

\begin{equation}\overline{BA+} = \cos(\theta) \label{AvgPlus}\end{equation}
and similarly

\begin{equation}\overline{BA-} = -\cos(\theta) \label{AvgMinus}\end{equation}
Putting these into Eq. (\ref{consCorrel}) we obtain
\begin{equation}\langle \alpha,\beta \rangle = \frac{1}{2}(+1)_A(\cos(\theta)) + \frac{1}{2}(-1)_A(-\cos(\theta)) = \cos(\theta) \label{consCorrel2}\end{equation}
which is the same as the quantum correlation function for the planar $S = 1$ conservation of spin angular momentum that we found above. Thus, we have an analogous picture for the ``SO(3) conservation'' of spin angular momentum for the $S = 1$ states as we had for the $S = 0$ state. Again, we point out that it is simply a mathematical fact that this ``average-only'' conservation yields the quantum correlation function. And again, Bob could partition the data according to his equivalence relation (per his reference frame) and claim that it is Alice who must average her results (obtained in her reference frame) to conserve spin angular momentum.

This all seems rather straightforward, the quantum correlation function for the Mermin device differs from that of instruction sets (classical correlation function) as necessary to satisfy conservation of spin angular momentum on average. And, the reason our conservation principle can only hold on average in different reference frames is because Alice and Bob only measure $\pm 1 \left(\frac{\hslash}{2}\right)$ (quantum), never a fraction of that amount (classical), as shown in Figure \ref{SGExp}. Indeed, many physicists are content with this explanation of Facts 1 and 2 for the Mermin device. But, stopping here would ignore what is clearly a conundrum for many other physicists. Therefore, we now articulate why there is still a ``mystery'' and how we propose to resolve it.

\section*{Conservation per No Preferred Reference Frame} \label{SecNPRF}
\begin{figure}[h]
\begin{center}
\includegraphics [width=\textwidth]{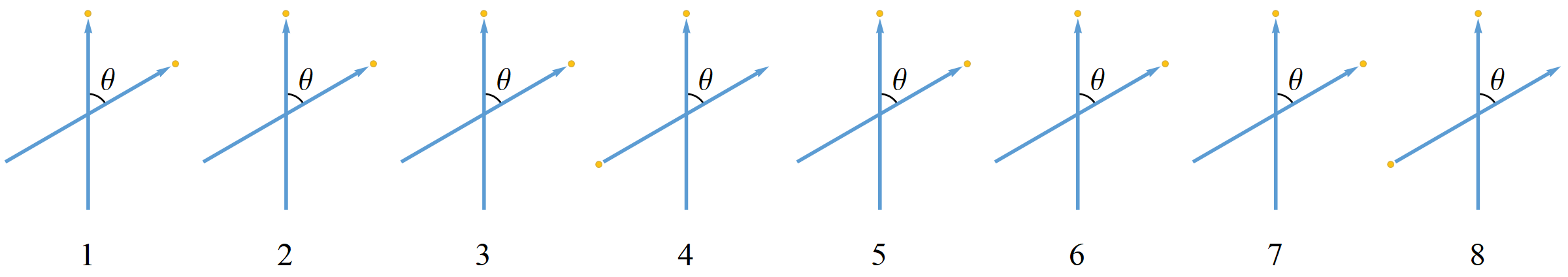} 
\caption{A spatiotemporal ensemble of 8 experimental trials for the spin triplet states showing Bob’s outcomes corresponding to Alice's $+1$ outcomes when $\theta = 60^\circ$. Spin angular momentum is not conserved in any given trial, because there are two different measurements being made, i.e., outcomes are in two different reference frames, but it is conserved on average for all 8 trials (six up outcomes and two down outcomes average to $\cos{(60^\circ)}=\frac{1}{2}$). It is impossible for spin angular momentum to be conserved explicitly in any given trial since the measurement outcomes are binary (quantum) with values of $+1$ (up) or $-1$ (down) per no preferred reference frame and explicit conservation of spin angular momentum in different reference frames would require a fractional outcome for Alice and/or Bob. The ``SO(3) conservation'' principle at work here does not assume Alice and Bob's measured values of angular momentum are mere components of some hidden spin angular momentum (Figures \ref{Projection1} \& \ref{Projection}). That is, the measured values of spin angular momentum \textit{are} the spin angular momenta contributing to this ``SO(3) conservation'' in accord with Dakic and Brukner's axiomatic reconstruction of quantum theory \cite{dakic}.} \label{4Dpattern}
\end{center}
\end{figure}

The problem with the average conservation principle responsible for the quantum correlation function is that it holds \textit{only on average} in different reference frames. Thus, it does not supply an explanation for outcomes on a trial-by-trial basis in different reference frames (Figure \ref{4Dpattern}). This is quite unlike constraints we have in classical physics. For example, conservation of momentum holds on a trial-by-trial basis because the sum of the forces equals zero and a light ray always takes the path of least time (Fermat's principle) because of refraction at the interface per Snell's law. Those constraints hold on average because they hold for each and every trial. In other words, constraints are often explained dynamically via causal mechanisms that hold on a trial-by-trial basis. Therefore in order to answer Mermin's challenge, we seek something other than a dynamical/causal mechanism to account for this ``average-only'' conservation in different reference frames, i.e., we seek a compelling principle. Essentially, we are in a situation with quantum mechanics that Einstein found himself in with special relativity \cite[pp. 51-52]{einstein1949}:
\begin{quote}
By and by I despaired of the possibility of discovering the true laws by means of constructive efforts based on known facts. The longer and the more despairingly I tried, the more I came to the conviction that only the discovery of a universal formal principle could lead us to assured results.
\end{quote}
That is, ``there is no mention in relativity of exactly \textit{how} clocks slow, or \textit{why} meter sticks shrink'' (no ``constructive efforts''), nonetheless the principles of special relativity are so compelling that ``physicists always seem so sure about the particular theory of Special Relativity, when so many others have been superseded in the meantime'' \cite{mainwood2018}. 

The principle we offer to explain ``average-only'' conservation in different reference frames is ``no preferred reference frame'' (NPRF), since it follows from the empirical facts. First, Bob and Alice both measure $\pm 1 \left(\frac{\hslash}{2}\right)$ for all SG magnet orientations relative to the source, i.e., relative ``to the vertical in the [symmetry] plane perpendicular to the line of flight of the particles.'' In order to satisfy conservation of spin angular momentum for any given trial when Alice and Bob are making different SG measurements in the symmetry plane, i.e., when they are in different reference frames, it would be necessary for Bob or Alice to measure some fraction, $\pm \cos(\theta)$, as we explained above. For example, if Alice measured $+1$ at $\alpha = 0$ for an $S = 1$ state (in the plane of symmetry) and Bob made his measurement (in the plane of symmetry) at $\beta = 60^\circ$, then Bob's outcome would need to be $\frac{1}{2}$ (Figure \ref{4Dpattern}). In that case, we would know that Alice measured the ``true'' spin angular momentum of her particle while Bob only measured a component of the ``true'' spin angular momentum for his particle. Thus, Alice's SG magnet orientation would definitely constitute a ``preferred reference frame.'' 

But, this is precisely what does \textit{not} happen. Alice and Bob both always measure $\pm 1 \left(\frac{\hslash}{2}\right)$, no fractions, in accord with NPRF. And, this fact alone distinguishes the quantum joint distribution from the classical joint distribution \cite{garg} (Figure \ref{SGExp}), so this fact alone also accounts for the elliptope constraint of Janas et al. Therefore, the ``average-only'' conservation responsible for the correlation function for the Bell spin states leading to Facts 1 and 2 for the Mermin device is actually conservation resulting from NPRF. Again, this is not the only counterintuitive result of NPRF in modern physics.

In special relativity, Alice is moving at velocity $\vec{V}_a$ relative to a light source and measures the speed of light from that source to be \textit{c} ($=\frac{1}{\sqrt{\mu_o\epsilon_o}}$, as predicted by Maxwell's equations). Bob is moving at velocity $\vec{V}_b$ relative to that same light source and measures the speed of light from that source to be \textit{c}. Here ``reference frame'' refers to the relative motion of the observer and source, so all observers who share the same relative velocity with respect to the source occupy the same reference frame. The corresponding transformation here is a Lorentz boost, which with our SO(3) transformation supra form the restricted Lorentz group. NPRF in this context thus means all measurements produce the same outcome \textit{c}. 

As a consequence of this constraint we have time dilation and length contraction, which are then reconciled per NPRF via the relativity of simultaneity. That is, Alice and Bob each partition spacetime per their own equivalence relations (per their own reference frames), so that equivalence classes are their own surfaces of simultaneity. If Alice's equivalence relation over the spacetime events yields the ``true'' partition of spacetime, then Bob must correct his lengths and times per length contraction and time dilation. Of course, the relativity of simultaneity says that Bob's equivalence relation is as valid as Alice's per NPRF. 

This is completely analogous to quantum mechanics, where Alice and Bob each partition the data per their own equivalence relations (per their own reference frames), so that equivalence classes are their own $+1$ and $-1$ data events. If Alice's equivalence relation over the data events yields the ``true'' partition of the data, then Bob must correct (average) his results per ``average-only'' conservation. Of course, NPRF says that Bob's equivalence relation is as valid as Alice's, which we might call the ``relativity of data partition'' (Table \ref{tab:SRvsQM}).

Thus, the counterintuitive aspects of special relativity (time dilation and length contraction) ultimately follow from the same principle as Mermin's ``Quantum mysteries for anybody,'' i.e., no preferred reference frame. Loosely speaking, NPRF is a ``unifying principle'' for non-relativistic quantum mechanics and special relativity per the restricted Lorentz symmetry group.

\section*{Discussion} \label{Secdiscussion}
As physicists work towards ``building a picture of the way nature works'' we are occasionally confronted with conundrums like that of quantum entanglement as conveyed by Mermin's challenge. Advancing physics calls for discharging such ``mysteries'' in order to ``say what is really going on out there.'' Weinberg states \cite[p. 5]{weinberg2017}:
\begin{quote}
What then must be done about the shortcomings of quantum mechanics? One reasonable response is contained in the legendary advice to inquiring students: ``Shut up and calculate!'' There is no argument about how to use quantum mechanics, only how to describe what it means, so perhaps the problem is merely one of words. On the other hand, the problems of understanding measurement in the present form of quantum mechanics may be warning us that the theory needs modification.
\end{quote}
That is, based on its ``shortcomings'' Weinberg suspects that quantum mechanics might actually require modification. Concerning this Mermin writes \cite[p. 2]{mermin2019}, ``Such modifications are motivated not by failures of the existing theory, but by philosophical discomfort with one or another of the prevailing interpretations of that theory.'' We agree with Mermin ``that if and when quantum mechanics is successfully modified, the motivation will come from unambiguous deviations of actual data from its predictions, and not from discomfort with any interpretations of its formalism'' \cite[p.2]{mermin2019}. 

In 2019, Mermin supplied his answer to what quantum mechanics means via his take on QBism \cite{mermin2019}. Accordingly \cite[p. 4]{mermin2019}: 
\begin{quote}
Laws of science are the regularities we have discerned in our individual experiences, and agreed on as a result of our communications with each other. Science, in general, and quantum mechanics, in particular, impose further constraints on my probabilistic expectations. They help each of us place better bets on our subsequent experience, based on our earlier experience.
\end{quote}
In other words, we (each of us) act on the world and the world responds. Quantum mechanics is telling each of us ``nothing more than the probability of the response I can expect'' \cite[p. 7]{mermin2019}. Of course, most physicists don't want to stop with this subjective account alone, since this still does not ``say what is really going on out there,'' i.e., it does not provide a corresponding objective account. Per QBism, our model of objective reality, i.e., our ``picture of the way nature works,'' is a collection of regularities/laws/constraints on individual experience, nothing more. So, QBism alone does not address the issue of actually constructing a model of objective reality and in that sense it does not actually address Mermin's challenge. However, as it turns out, our ``conservation per NPRF'' answer to Mermin's challenge is perfectly compatible with Mermin's take on QBism and also addresses Weinberg's issue with quantum mechanics without modification to quantum mechanics. 
Weinberg writes \cite[p. 3]{weinberg2017}:
\begin{quote}
An electron spin that has not been measured is like a musical chord, formed from a superposition of two notes that correspond to positive or negative spins, each note with its own amplitude. Just as a chord creates a sound distinct from each of its constituent notes, the state of an electron spin that has not yet been measured is a superposition of the two possible states of definite spin, the superposition differing qualitatively from either state. In this musical analogy, the act of measuring the spin somehow shifts all the intensity of the chord to one of the notes, which we then hear on its own. ...\\

So if we regard the whole process of measurement as being governed by the equations of quantum mechanics, and these equations are perfectly deterministic, \textit{how do probabilities get into quantum mechanics}?
\end{quote}
His issue with quantum mechanics is that the deterministic quantum formalism in Hilbert space does not translate into deterministic measurement outcomes in real space. The reason for that as regards entangled spin measurements is our answer to Mermin's challenge, i.e., ``conservation per NPRF.'' [For our answer in general see our work here\cite{ourbook}.] As Weinberg notes, there are only two possible outcomes for the measurement of electron spin, ``One possible result will be equal to a positive number, a universal constant of nature. ... The other possible result is its opposite, the negative of the first'' \cite[p. 3]{weinberg2017}. That obtains because NPRF applies to the measurement of universal constants of nature, like $h$ and $c$. So, why is it possible for the deterministic state vector in Hilbert space to fall between the only two possible outcomes? In other words, why doesn't quantum mechanics just deterministically give us $\pm \frac{\hslash}{2}$? Again, the answer to that question in the present context is absolutely clear. The Hilbert space representation of the entangled quantum state (Bell spin state) is giving us the distribution of correlated $\pm \frac{\hslash}{2}$ outcomes such that spin angular momentum is conserved on average between different reference frames with Alice and Bob each measuring $+ \frac{\hslash}{2}$ and $- \frac{\hslash}{2}$ with equal frequency in all reference frames. And, ``on average'' is the only way spin angular momentum can be conserved between different reference frames, since there are only two possible outcomes. Of course, both $+ \frac{\hslash}{2}$ and $- \frac{\hslash}{2}$ have to be possible in order to be able to obtain the required fractional average. In short, the Bell spin states can be \textit{derived} from ``conservation per NPRF'' \cite{TsirelsonBound2019}.

According to ``conservation per NPRF,'' the deepest truth about ``what is really going on out there'' is that the regularities/laws/constraints on individual experience and their associated constants are accessible to anyone or any thing (full disclosure, no ``hidden variables'') such that no one or no thing has privileged access to them. Earth is not the center of the universe, there is no reference frame in which the speed of light is uniquely given by $c =\frac{1}{\sqrt{\mu_o\epsilon_o}}$, and there is no reference frame in which Planck's constant is uniquely \textit{h}. The consequences are often strongly counterintuitive, i.e., \textit{clearly} everything in the sky revolves around us, \textit{clearly} it should be possible to measure different values for the speed of light when moving relative to the source at different velocities, and \textit{clearly} Alice or Bob has to be able to measure some fraction of $\frac{\hslash}{2}$ in order to conserve spin angular momentum when making entangled spin measurements at different angles. What we showed herein is that when NPRF is applied to the measurement of Planck's constant in the context of entangled (conserved) spin angular momentum, the consequence is ``average-only'' conservation, i.e., probability that obtains deterministically and unavoidably. As Mermin states \cite[p. 10]{mermin2019}, ``Quantum mechanics is, after all, the first physical theory in which probability is explicitly not a way of dealing with ignorance of the precise values of existing quantities.'' And we see why that is in our answer to Mermin's challenge.

The use of symmetries to guide the progress of physics is already well established and symmetries are just another way of expressing constraints and conservation principles. The symmetry group relating non-relativistic quantum mechanics and special relativity via their ``mysteries'' as shown herein is the restricted Lorentz group. Again and again, symmetries have served to advance and unify physics. While NPRF has profoundly counterintuitive implications, it has not kept us from ``building a picture of the way nature works.'' On the contrary, given the enormous success of physics, the egalitarian transparency of nature seems to have facilitated our attempts to ``say what is really going on out there.'' All we have to do to appreciate the coherence and integrity of what we find is to discard our anthropocentric biases. After all, the human species is a part of nature and is therefore subject to its fundamental principles, so no preferred reference frame entails no anthropocentricism. And that has implications for ``the wider world of human activity outside of science.''

\begin{table}[]
    \centering
    \begin{tabular}{|l|l|}
    \hline
        {\bf Special Relativity} & {\bf Quantum Mechanics} \\
        \hline
        Empirical Fact: Alice and Bob both measure $c$,  & Empirical Fact: Alice and Bob both measure $\pm 1 \left (  \frac {\hslash}2 \right )$,   \\
     regardless of their motion relative to the source  &  regardless of their SG orientation relative to the source \\
        \hline
        Alice(Bob) says of Bob(Alice): Must correct time  &  Alice(Bob) says of Bob(Alice): Must average results \\ 
        and length measurements &   \\
        \hline
        NPRF: Relativity of simultaneity & NPRF: Relativity of data partition \\
        \hline
    \end{tabular}
    \caption{\textbf{Comparing special relativity with quantum mechanics according to no preferred reference frame (NPRF)}. Because Alice and Bob both measure the same speed of light $c$, regardless of their motion relative to the source per NPRF, Alice(Bob) may claim that Bob's(Alice's) length and time measurements are erroneous and need to be corrected (length contraction and time dilation). Likewise, because Alice and Bob both measure the same values for spin angular momentum $\pm 1$ $\left(\frac{\hslash}{2}\right)$, regardless of their SG magnet orientation relative to the source per NPRF, Alice(Bob) may claim that Bob's(Alice's) individual $\pm 1$ values are erroneous and need to be corrected (averaged, Figures \ref{AvgViewSinglet}, \ref{AvgViewTriplet}, \& \ref{4Dpattern}). In both cases, NPRF resolves the ``mystery'' it creates. In special relativity, the apparently inconsistent results can be reconciled via the relativity of simultaneity. That is, Alice and Bob each partition spacetime per their own equivalence relations (per their own reference frames), so that equivalence classes are their own surfaces of simultaneity and these partitions are equally valid per NPRF. This is completely analogous to quantum mechanics, where the apparently inconsistent results per the Bell spin states arising because of NPRF can be reconciled by NPRF via the ``relativity of data partition.'' That is, Alice and Bob each partition the data per their own equivalence relations (per their own reference frames), so that equivalence classes are their own $+1$ and $-1$ data events and these partitions are equally valid.}
    \label{tab:SRvsQM}
\end{table}

\section*{Methods} \label{Secmethods}
Here we provide the interested reader with the mathematical details justifying the results in The Bell Spin States. The Bell spin states of Eq. \ref{BellStates} are given in the eigenbasis of $\sigma_z$ where the Pauli spin matrices are
$$
\sigma_x = \left ( \begin{array}{rr} 0 & \phantom{00}1 \\ 1 & \phantom{0}0 \end{array} \right ), \quad
\sigma_y = \left ( \begin{array}{rr} 0 & \phantom{0}-i \\ i & 0  \end{array} \right ), \quad \mbox{and} \quad
\sigma_z =\left ( \begin{array}{rr} 1 & 0 \\ 0 & \phantom{0}-1 \end{array} \right ).
$$
All spin matrices have the same eigenvalues of $\pm 1$ and we will denote the corresponding eigenvectors as $|u\rangle$ and $|d\rangle$ for spin up ($+1$) and spin down ($-1$), respectively. Using the Pauli spin matrices above with $|u\rangle = \left ( \begin{array}{rr} 1 \\ 0\end{array} \right )$ and $|d\rangle = \left ( \begin{array}{rr} 0 \\ 1\end{array} \right )$, we see that $\sigma_z|u\rangle = |u\rangle$, $\sigma_z|d\rangle = -|d\rangle$, $\sigma_x|u\rangle = |d\rangle$, $\sigma_x|d\rangle = |u\rangle$, $\sigma_y|u\rangle = i|d\rangle$, and $\sigma_y|d\rangle = -i|u\rangle$. We will use the juxtaposed notation in Eq. (\ref{BellStates}) for our spin matrices as well. Thus, $\sigma_x\sigma_z|ud\rangle = -|dd\rangle$ and $\sigma_x\sigma_y|ud\rangle = -i|du\rangle$, for example. Essentially, this notation is simply ignoring the tensor product sign $\otimes$, so that $\left(\sigma_x \otimes \sigma_z \right)||u\rangle \otimes |d\rangle \rangle = \sigma_x\sigma_z|ud\rangle$. It will be obvious which spin matrix is acting on which Hilbert space vector via the juxtaposition. If we flip the orientation of a vector from right pointing (ket) to left pointing (bra) or vice-versa, we transpose and take the complex conjugate. For example, if $|A\rangle =i\begin{pmatrix} 1\\0 \end{pmatrix} = i|u\rangle$, then $\langle A| = -i\begin{pmatrix} 1\;\; 0 \end{pmatrix} = -i\langle u|$. Thus, any spin matrix can be written as $(+1)|u\rangle\langle u| + (-1)|d\rangle\langle d|$ where $|u\rangle$ and $|d\rangle$ are their up and down eigenvectors, respectively. With that review of the formalism, we now explore the conservation being depicted by the Bell spin states and relate it to the correlation function. Let us start with the spin singlet state $|\psi_-\rangle$.

If we transform our basis per
\begin{equation}
\begin{aligned}
|u\rangle &\rightarrow &\cos(\Theta)|u\rangle + \sin(\Theta)|d\rangle \\
|d\rangle &\rightarrow &-\sin(\Theta)|u\rangle + \cos(\Theta)|d\rangle \label{Xtransform} 
\end{aligned}
\end{equation}
where $\Theta$ is an angle in Hilbert space (as opposed to the SG magnet angles in real space), then $|\psi_-\rangle \rightarrow |\psi_-\rangle$. In other words, $|\psi_-\rangle$ is invariant with respect to this SU(2) transformation. Constructing the corresponding spin measurement operator from these transformed up and down vectors gives
\begin{equation} |u\rangle\langle u| - |d\rangle\langle d| = \left (\begin{array}{rr} \cos(2\Theta) & \sin(2\Theta)\\\sin(2\Theta) & -\cos(2\Theta) \end{array} \right ) = \cos(2\Theta)\sigma_z + \sin(2\Theta)\sigma_x \label{sigmaOp1} \end{equation}
So, we see that the invariance of the state under this Hilbert space SU(2) transformation means we have rotational (SO(3)) invariance for the SG measurement outcomes in the $xz$-plane of real space. Specifically, $|\psi_-\rangle$ says that when the SG magnets are aligned in the $z$ direction (Alice and Bob are in the same reference frame) the outcomes are always opposite ($\frac{1}{2}$ $ud$ and $\frac{1}{2}$ $du$). Since $|\psi_-\rangle$ has that same functional form under an SU(2) transformation in Hilbert space representing an SO(3) rotation in the $xz$-plane per Eqs. (\ref{Xtransform}) \& (\ref{sigmaOp1}), the outcomes are always opposite ($\frac{1}{2}$ $ud$ and $\frac{1}{2}$ $du$) for aligned SG magnets in the $xz$-plane. That is the ``SO(3) conservation'' associated with this SU(2) symmetry. Note that it only deals with case (a) results, i.e., when Alice and Bob are in the same reference frame, so this alone does not distinguish between the Mermin device and instruction sets. 

From Eq. (\ref{sigmaOp1}) we see that when the angle in Hilbert space is $\Theta$, the angle $\theta$ of the rotated SG magnets in the $xz$-plane is $\theta = 2\Theta$. The physical reason for this factor of $2$ relating $\Theta$ in Hilbert space and $\theta$ in real space was made evident above when we revealed the implications of the ``SO(3) conservation'' for measurements in different reference frames (Figures \ref{AvgViewSinglet} \& \ref{AvgViewTriplet}). Notice that when $\Theta = 45^\circ$, our operator is $\sigma_x$, i.e., we have transformed to the eigenbasis of $\sigma_x$ from the eigenbasis of $\sigma_z$.

Another SU(2) transformation that leaves $|\psi_-\rangle$ invariant is

\begin{equation}
\begin{aligned}
&|u\rangle \rightarrow &\cos(\Theta)|u\rangle + i\sin(\Theta)|d\rangle \\ 
&|d\rangle \rightarrow &i\sin(\Theta)|u\rangle + \cos(\Theta)|d\rangle \label{Ytransform} 
\end{aligned}
\end{equation}
Constructing our spin measurement operator from these transformed vectors gives us

\begin{equation} |u\rangle\langle u| - |d\rangle\langle d| = \left ( \begin{array}{rr} \cos(\theta) & -i\sin(\theta)\\i\sin(\theta) & -\cos(\theta) \end{array} \right ) = \cos(\theta)\sigma_z + \sin(\theta)\sigma_y \end{equation}
So, we see that the invariance of the state under this Hilbert space SU(2) transformation means we have rotational (SO(3)) invariance for the SG measurement outcomes in the $yz$-plane, analogous to what we found for the $xz$-plane. Notice that when $\Theta = 45^\circ$ our operator is $\sigma_y$, i.e., we have transformed to the eigenbasis of $\sigma_y$ from the eigenbasis of $\sigma_z$.  

Finally, we see that $|\psi_-\rangle$ is invariant under the third SU(2) transformation

\begin{equation}
\begin{aligned}
&|u\rangle \rightarrow &(\cos(\Theta) + i\sin(\Theta))|u\rangle \\ 
&|d\rangle \rightarrow &(\cos(\Theta) - i\sin(\Theta))|d\rangle \label{Ztransform} 
\end{aligned}
\end{equation}
since this takes $|ud\rangle \rightarrow |ud\rangle$. Constructing our spin measurement operator from these transformed vectors gives us

\begin{equation} |u\rangle\langle u| - |d\rangle\langle d| = \left ( \begin{array}{rr} 1 & 0\\0 & -1 \end{array} \right ) = \sigma_z \end{equation}
In other words, Eq. (\ref{Xtransform}) is the Hilbert space SU(2) transformation that represents an SO(3) rotation about the $y$ axis in real space and can be written
\begin{equation}
    \left ( \begin{array}{rr} u \\ d\end{array} \right ) \rightarrow \left (\begin{array}{rr} \cos(\Theta) & \sin(\Theta)\\-\sin(\Theta) & \cos(\Theta) \end{array} \right )\left ( \begin{array}{rr} u \\ d\end{array} \right ) = \left( \cos(\Theta)I + i\sin(\Theta)\sigma_y \right)\left ( \begin{array}{rr} u \\ d\end{array} \right )
\end{equation}
Eq. (\ref{Ytransform}) is the Hilbert space SU(2) transformation that represents an SO(3) rotation about the $x$ axis in real space and can be written
\begin{equation}
    \left ( \begin{array}{rr} u \\ d\end{array} \right ) \rightarrow \left (\begin{array}{rr} \cos(\Theta) & i\sin(\Theta)\\i\sin(\Theta) & \cos(\Theta) \end{array} \right )\left ( \begin{array}{rr} u \\ d\end{array} \right ) = \left( \cos(\Theta)I + i\sin(\Theta)\sigma_x \right)\left ( \begin{array}{rr} u \\ d\end{array} \right )
\end{equation}
And Eq. (\ref{Ztransform}) is the Hilbert space SU(2) transformation that represents an SO(3) rotation about the $z$ axis in real space and can be written
\begin{multline}
    \left ( \begin{array}{rr} u \\ d\end{array} \right ) \rightarrow \left (\begin{array}{cc} \cos(\Theta) + i\sin(\Theta) & 0\\0 & \cos(\Theta) -i\sin(\Theta) \end{array} \right )\left ( \begin{array}{rr} u \\ d\end{array} \right )  =
    \left( \cos(\Theta)I + i\sin(\Theta)\sigma_z \right)\left ( \begin{array}{rr} u \\ d\end{array} \right )
\end{multline}
The SU(2) transformation matrix is often written $e^{i\Theta\sigma_j}$, where $j = \{x,y,z\}$, by expanding the exponential and using $\sigma_j^2 = I$. Since we are in the $\sigma_z$ eigenbasis, this third transformation means our spin measurement operator is just $\sigma_z$. The invariance of $|\psi_-\rangle$ under all three SU(2) transformations makes sense, since the spin singlet state represents the conservation of a total spin angular momentum of $S = 0$, which is directionless, and each SU(2) transformation in Hilbert space corresponds to an element of SO(3) in real space.

So, while we know that invariance under this third SU(2) transformation means we have rotational (SO(3)) invariance of our SG measurement outcomes in the $xy$-plane, we do not know what those outcomes are unless we rotate our state to one of those eigenbases. That is, we need to know what this state says about the SG measurement outcomes when the SG magnets are aligned in the $xy$-plane. Since $|\psi_-\rangle$ is invariant under either of the other SU(2) transformations, it has the same form in either the $\sigma_x$ or $\sigma_y$ eigenbasis. Thus, the SG measurement outcomes are always opposite ($\frac{1}{2}$ $ud$ and $\frac{1}{2}$ $du$) for aligned SG magnets in any plane of real space. This will not be the case for the spin triplet state $|\psi_+\rangle$ that is invariant under this third SU(2) transformation, as it is \textit{only} invariant under this third SU(2) transformation.

Now, since our state has the same functional form in any plane, we are free to choose any plane we like to compute our correlation function and not lose generality. Let us work in the eigenbasis of $\sigma_1 = \sigma_z$ with $\sigma_2 = \cos(\theta)\sigma_z + \sin(\theta)\sigma_x$ in computing our correlation function for $|\psi_-\rangle$. We have

\begin{equation} \frac{1}{2}(\langle ud| - \langle du|)\sigma_z [\cos(\theta)\sigma_z + \sin(\theta)\sigma_x](|ud\rangle - |du\rangle) = -\cos(\theta) \label{PsiMinuscorr} \end{equation}
per the rules of the formalism in agreement with Eq. (\ref{gencorrelations}), which gives $-\hat{a}\cdot\hat{b}$. What we see from this analysis is that the conserved spin angular momentum ($S = 0$), being directionless, leads to opposite outcomes for SG magnets at any $\hat{a}=\hat{b}$ and a correlation function of $-\cos(\theta)$ in any plane of real space. As we saw above, this correlation function tells us there are case (b) implications for our case (a) conservation. Now for the spin triplet states.

We will begin with $|\phi_+\rangle$. The only SU(2) transformation that takes $|\phi_+\rangle \rightarrow |\phi_+\rangle$ is Eq. (\ref{Xtransform}). Thus, this state says we have rotational (SO(3)) invariance for our SG measurement outcomes in the $xz$-plane. Specifically, $|\phi_+\rangle$ says that when the SG magnets are aligned in the $z$ direction (measurements are being made in the same reference frame) the outcomes are always the same ($\frac{1}{2}$ $uu$ and $\frac{1}{2}$ $dd$). Since $|\psi_+\rangle$ has that same functional form under an SU(2) transformation in Hilbert space representing an SO(3) rotation in the $xz$-plane per Eqs. (\ref{Xtransform}) \& (\ref{sigmaOp1}), the outcomes are always the same ($\frac{1}{2}$ $uu$ and $\frac{1}{2}$ $dd$) for aligned SG magnets in the $xz$-plane. Again, that is the ``SO(3) conservation'' associated with this SU(2) symmetry and it applies only to case (a), i.e., measurements made in the same reference frame. In this case, since $|\phi_+\rangle$ is only invariant under Eq. (\ref{Xtransform}), we can only expect rotational invariance for our SG measurement outcomes in the $xz$-plane. This is confirmed by Eq. (\ref{gencorrelations}) where we see that the correlation function for arbitrarily oriented $\sigma_1$ and $\sigma_2$ is given by $a_xb_x - a_yb_y + a_zb_z$. Thus, unless we restrict our measurements to the $xz$-plane, we do not have the rotationally invariant correlation function $\hat{a}\cdot\hat{b}$ analogous to the spin singlet state. Restricting our measurements to the $xz$-plane gives us

\begin{equation} 
\frac{1}{2}(\langle uu| + \langle dd|)\sigma_z [\cos(\theta)\sigma_z + \sin(\theta)\sigma_x](|uu\rangle + |dd\rangle) = \cos(\theta) \label{PhiPluscorr} \end{equation}
per the rules of the formalism in agreement with Eq. (\ref{gencorrelations}). Again, as we saw above, this correlation function tells us there are case (b) implications for our case (a) conservation. We next consider $|\phi_-\rangle$.

The only SU(2) transformation that leaves $|\phi_-\rangle$ invariant is Eq. (\ref{Ytransform}). Thus, this state says we have rotational (SO(3)) invariance for the SG measurement outcomes in the $yz$-plane. Since $|\phi_-\rangle$ is only invariant under Eq. (\ref{Ytransform}), we can only expect rotational invariance for our SG measurement outcomes in the $yz$-plane. This is confirmed by Eq. (\ref{gencorrelations}) where we see that the correlation function for arbitrarily oriented $\sigma_1$ and $\sigma_2$ for $|\phi_-\rangle$ is given by $-a_xb_x + a_yb_y + a_zb_z$. Thus, unless we restrict our measurements to the $yz$-plane, we do not have the rotationally invariant correlation function $\hat{a}\cdot\hat{b}$ analogous to the spin singlet state. Restricting our measurements to the $yz$-plane gives us

\begin{equation} \frac{1}{2}(\langle uu| - \langle dd|)\sigma_z [\cos(\theta)\sigma_z + \sin(\theta)\sigma_y](|uu\rangle - |dd\rangle) = \cos(\theta) \label{PhiMinuscorr} 
\end{equation}
per the rules of the formalism in agreement with Eq. (\ref{gencorrelations}). 

Finally, the only SU(2) transformation that leaves $|\psi_+\rangle$ invariant is Eq. (\ref{Ztransform}). Thus, this state says we have rotational (SO(3)) invariance for our SG measurement outcomes in the $xy$-plane. But, unlike the situation with $|\psi_-\rangle$, we will need to transform to either the $\sigma_x$ or $\sigma_y$ eigenbasis to see what we are going to find in the $xy$-plane. We can either transform first from the $\sigma_z$ eigenbasis to the $\sigma_x$ eigenbasis and then look for our SU(2) invariance transformation, or first transform from the $\sigma_z$ eigenbasis to the $\sigma_y$ eigenbasis. We will do $\sigma_z$ to $\sigma_x$, the other is similar ($|\psi_+\rangle$ in the $\sigma_z$ eigenbasis goes to $i|\phi_+\rangle$ in the $\sigma_y$ eigenbasis and we know the transformation that leaves this invariant is Eq. (\ref{Xtransform})).

To go to the $\sigma_x$ eigenbasis from the $\sigma_z$ eigenbasis we use Eq. (\ref{Xtransform}) with $\Theta = 45^\circ$

\begin{equation}|u\rangle \rightarrow \frac{1}{\sqrt{2}}|u\rangle + \frac{1}{\sqrt{2}}|d\rangle \end{equation}
\begin{equation}|d\rangle \rightarrow -\frac{1}{\sqrt{2}}|u\rangle + \frac{1}{\sqrt{2}}|d\rangle \end{equation}
This takes $|\psi_+\rangle$ in the $\sigma_z$ eigenbasis to $-|\phi_-\rangle$ in the $\sigma_x$ eigenbasis and we know the SU(2) transformation that leaves this invariant is Eq. (\ref{Ytransform}) which then gives a spin measurement operator of $\cos(\theta)\sigma_x + \sin(\theta)\sigma_y$, since we have simply switched the $\sigma_z$ eigenbasis with the $\sigma_x$ eigenbasis. Therefore, $|\psi_+\rangle$ says that when the SG magnets are aligned anywhere in the $xy$-plane the outcomes are always the same ($\frac{1}{2}$ $uu$ and $\frac{1}{2}$ $dd$). This is consistent with Eq. (\ref{gencorrelations}) where we see that the correlation function for arbitrarily oriented $\sigma_1$ and $\sigma_2$ for $|\psi_+\rangle$ is given by $a_xb_x + a_yb_y - a_zb_z$. Thus, unless we restrict our measurements to the $xy$-plane, we do not have the rotationally invariant correlation function $\hat{a}\cdot\hat{b}$ analogous to the spin singlet state. Restricting our measurements to the $xy$-plane gives us

\begin{equation} \frac{1}{2}(\langle uu| - \langle dd|)\sigma_x [\cos(\theta)\sigma_x + \sin(\theta)\sigma_y](|uu\rangle - |dd\rangle) = \cos(\theta) \label{PsiPluscorr} 
\end{equation}
where $|u\rangle$ and $|d\rangle$ are now the eigenstates for $\sigma_x$. That is, $|u\rangle = \left ( \begin{array}{rr} 1/\sqrt{2} \\ 1/\sqrt{2}\end{array} \right )$ and $|d\rangle = \left ( \begin{array}{rr} -1/\sqrt{2} \\ 1/\sqrt{2}\end{array} \right )$, so that $\sigma_x|u\rangle = |u\rangle$, $\sigma_x|d\rangle = -|d\rangle$, $\sigma_y|u\rangle = i|d\rangle$, and $\sigma_y|d\rangle = -i|u\rangle$. Again, this agrees with Eq. (\ref{gencorrelations}). 

The reader interested in how conservation per NPRF relates to the more general Clauser-Horne-Shimony-Holt (CHSH) inequality, the quantum states proper, the Tsirelson bound, and the Malus law may read our work here\cite{TsirelsonBound2019}. In addition to the analogy with special relativity mentioned in The Bell Spin States, an anonymous reviewer points out the following. Galilean boosts commute and are obtained from $c \rightarrow \infty$ in the Lorentz boosts which do not commute. In quantum mechanics, the position and momentum operators do not commute, i.e., they are said to be ``complementary.'' In classical mechanics, position and momentum operators commute and that commutation relation is obtained from $h \rightarrow 0$ in the commutation relations for position and momentum operators in quantum mechanics. Thus, the fact that everyone must measure the same value $c$ for the speed of light regardless of their velocity relative to the source means different reference frames in special relativity are ``complementary'' in the language of quantum mechanics. Of course, the Pauli spin operators which represent different reference frames for SG spin measurements do not commute and so they are also complementary. Just as with position and momentum operators of quantum mechanics, the Pauli spin operators would commute if $h \rightarrow 0$.

Finally, since $\sigma_1$ and $\sigma_2$ establish frames of reference one might say that $\Sigma_{same} := \sigma_1\sigma_2$ for $\hat{a} = \hat{b}$ constitutes a preferred reference frame in violation of NPRF in the sense that we obtain exact conservation in the relevant symmetry plane for $\Sigma_{same}$ while we obtain ``average-only'' conservation for $\Sigma_{diff} := \sigma_1\sigma_2$ for $\hat{a} \ne \hat{b}$.  In terms of Hilbert space, we are saying that the dynamical evolution of the Bell spin states ($|n\rangle, n = 1,2,3,4$) under $\Sigma_{same}$ is different than $\Sigma_{diff}$, since $\langle n'|\Sigma_{same}|n \rangle = 0, n' \ne n$ while $\langle n'|\Sigma_{diff}|n \rangle \ne 0, n' \ne n$ because the Bell spin states are eigenstates of $\Sigma_{same}$, while not of $\Sigma_{diff}$. But far from violating NPRF, this situation obtains \textit{because} of NPRF. As we pointed out in the Discussion, quantum mechanics is necessarily probabilistic in this situation because of conservation per NPRF. The difference being pointed out is simply a difference in the degree of that probability. As $\alpha$ deviates more and more from $\beta$, the average conservation deviates more and more from the exact conservation that obtains for $\alpha = \beta$, where exact conservation can be viewed as the ``probability 1'' case. For example, consider measurements of a spin triplet state as depicted in Figures \ref{AvgViewTriplet} \& \ref{4Dpattern}. For Alice's $+1$ results at $\theta = 0$, Bob's results must average to $+1$. That means his distribution of $+1$ and $-1$ results is exclusively $+1$, i.e., the probability of him measuring $+1$ is 1. As $\theta$ increases, his distribution of $+1$ and $-1$ results gradually acquires more $-1$ data points, so that the probability of him measuring $+1$ diminishes. When $\theta = \frac{\pi}{2}$, Bob is measuring equal numbers of $+1$ and $-1$ results, so the probability of him measuring $+1$ has reduced to $\frac{1}{2}$. When $\theta = \pi$, Bob is measuring exclusively $-1$ results, so the probability of him measuring $+1$ has reduced to 0. Thus, the $\theta = 0$ case ($\Sigma_{same}$ case) can be understood simply as residing on one end of a probabilistic continuum in accord with conservation per NPRF.

\bibliography{biblio}

\section*{Author contributions statement}
W.S. physics and writing, M.S. conceptual analysis and writing, T.M. math and figures, T.L. physics. All authors reviewed the manuscript. 

\section*{Additional information}
\textbf{Competing interests}.
The authors declare no competing interests.

\end{document}